\documentclass[conference]{IEEEtran}
\usepackage{multicol}
\usepackage{amsfonts}
\makeatletter
 \def\ps@headings{%
 \def\@oddhead{\mbox{}\scriptsize\rightmark \hfil \thepage}%
 \def\@evenhead{\scriptsize\thepage \hfil \leftmark\mbox{}}%
 \def\@oddfoot{}%
 \def\@evenfoot{}}
 \makeatother
\usepackage{times,amsmath,color,amssymb,graphicx,epsfig,cite,psfrag,subfigure,lscape,array}
\usepackage{times,amsmath,color,amssymb,graphicx,epsfig,cite,psfrag,subfigure,algorithm,algorithmic,multirow}

\date{}

\newcommand{\mbf}{\boldmath}

\def\b0{{\mbox {\mbf $0$}}}

\def\bb0{{\mathbf{0}}}

\def\b_beta{\mbox{\boldmath $\beta$}}
\def\hb_beta{\mbox{\boldmath $\hat \beta$}}

\IEEEoverridecommandlockouts
\begin{document}

\title{OTFS for Joint Radar and Communication: Algorithms, Prototypes, and Experiments}

\author{\IEEEauthorblockN{Xiaojuan Zhang, Yonghong Zeng, Francois Chin Po Shin}
   Institute for Infocomm Research, A*STAR, Singapore 138632
     \thanks{This research is supported by the National Research Foundation, Singapore and Infocomm Media Development Authority under its Future Communications Research Development Programme.}} \maketitle
\maketitle

\begin{abstract}
We propose an Joint Radar and Communication (JRC) system that utilizes the Orthogonal Time Frequency Space (OTFS) signals. The system features a fast radar sensing algorithm for detecting target range and speed by using the OTFS communication signals, and a self-interference cancellation for enhanced multi-target separation.  In addition to target detection, we propose methods for monitoring human vital signs, such as breathing rate and heartbeat. Furthermore, we explore two approaches for distinguishing between human and non-human targets: one based on signal processing and the other based on machine learning. We have developed a prototype JRC system using the software-defined radio (SDR) technology. Experimental results are shown to demonstrate the effectiveness of the prototype in detecting range, speed, and vital signs in both human and mobile robot scenarios, as well as in distinguishing between human and non-human targets. 
\end{abstract}

\section{Introduction}

In November 2023, the ITU formally incorporated integrated sensing and communication (ISAC) into its IMT-2030 framework. Joint Radar and Communication (JRC) is a specific type of ISAC that utilizes shared hardware and waveforms to efficiently perform both sensing and communication tasks. This approach offers significant benefits in spectrum utilization, power consumption, and overall costs, with wide-ranging applications in areas such as assisted and automated driving systems \cite{IEEE:ISC} -\cite{IEEE:SIS}.

Orthogonal Time Frequency Space (OTFS) modulation is an innovative scheme designed to tackle the challenges of wireless communication in fast-moving environments. Traditional modulation techniques often struggle with reliability and efficiency under high mobility conditions, such as in vehicular communication. OTFS addresses these issues by representing signals in a joint time and frequency domain, effectively decoupling time and frequency. This unique approach mitigates the adverse effects of Doppler shifts and delay spreads, making OTFS particularly effective in dynamic environments. Consequently, integrating OTFS into JRC systems is a promising area of research within the context of 6G technology \cite{IEEE:PAJ}-\cite{IEEE:DDC}.

In this paper, we propose a fast radar sensing algorithm for detecting target range and speed in JRC systems using OTFS signals, with the integration of self-interference cancellation techniques for enhanced multi-target separation. Additionally, we propose methods for detecting human vital signs, such as breathing rate and heartbeat, following the initial target detection. Furthermore, we explore two approaches to differentiate between human and non-human targets: one based on signal processing and the other based on machine learning.

While many theoretical models for JRC are based on idealized assumptions, real-world environments introduce challenges such as noise, interference, and hardware limitations. Experimental studies on JRC, especially those utilizing OTFS signals, remain limited. To address this gap, we developed a JRC prototype to evaluate the practical performance of our algorithms. Our experimental results demonstrate the prototype's effectiveness in detecting range, speed, and vital signs in both human and moving robot scenarios, as well as distinguishing between human and non-human targets. These findings validate theoretical models, reveal environmental and hardware constraints, and provide valuable insights for refining algorithms, enhancing real-world applications, and assessing commercial feasibility.



\section{System Model}

\begin{figure}[h!]

        \centering
        \includegraphics[width=8cm,angle=0]{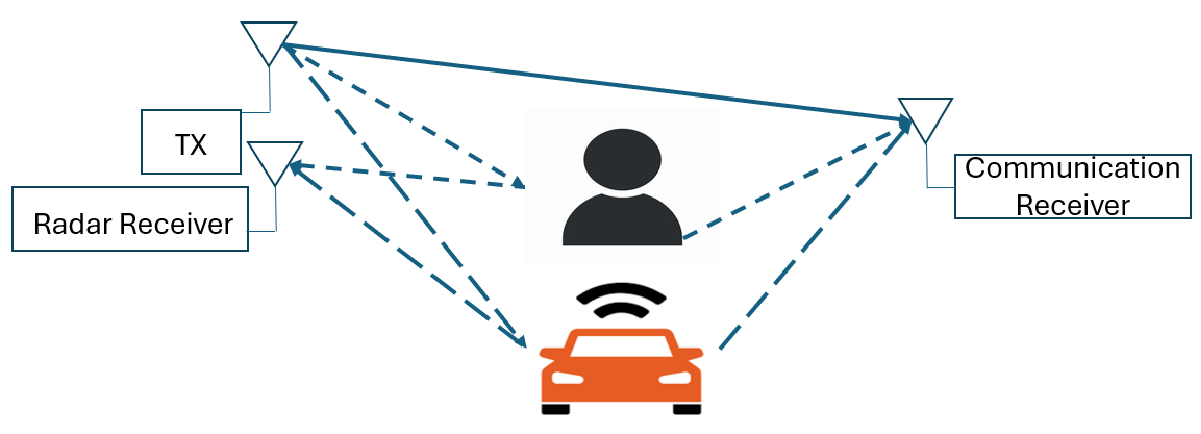}

\vspace{-2mm}
        \caption{An OTFS JRC system with multiple targets}
        \label{fig:model}
\end{figure}

The system model for the OTFS Joint Radar and Communication (JRC) system is depicted in Fig.~\ref{fig:model}. In this setup, the transmitter (TX) produces an OTFS communication signal, which is sent to the target through a TX antenna (ANT). The communication receiver is responsible for signal demodulation and decoding, while the radar receiver is dedicated to radar detection. This paper primarily focuses on the radar detection aspect. By capturing and analyzing the reflected signals from radar targets, the radar receiver estimates both the range (distance) and velocity of these targets. If the detected radar target is a human, the system can also estimate vital signs, including breathing rate and heartbeat. While Fig.~\ref{fig:model} presents two targets for simplicity, the system is engineered to detect and differentiate multiple targets based on their range and velocity, demonstrating its effectiveness in complex environments with numerous targets.

In OTFS modulation, the transmitted data $X(m,n)$ $(m=0,1,\cdots,M-1; \ n=0,1,\cdots,N-1)$ is placed in the delay-Doppler (DD) domain \cite{IEEE:PAJ}-\cite{IEEE:DDC}.  The DD domain signal $X(m, n)$ is converted to the delay-time domain signal through the row inverse fast Fourier transform (IFFT) and transmitted. At the receiver, the received signal passed through the channel is transformed by the row FFT to the DD domain signal $Y(m,n)$. We assume that there are $P$ radar targets. Let $\Delta_f$ be the subcarrier spacing. Then $M\Delta_f$ is the sampling rate of the OTFS signal, and $T_s=1/(M\Delta_f)$ is the sampling period. The DD domain received signal can be approximated as \cite{IEEE:ODDD,IEEE:IDOT}
\begin{align}
&Y(m,n)=\sum_{p=1}^P h_p e^ {j 2\pi \left(\frac{m-l_p}{M}\right) \frac{k_p}{N}} \alpha_p (m,n)\nonumber\\
&\cdot X\left(\left< m-l_p\right>_m,\left< n-k_p\right>_N \right) + \eta(m,n)\label{eq-OTFS1}
\end{align}
where $l_p=\tau_p/T_s$ ($0\leq l_p\leq M-1$) is the normalized delay for the $p$-th target ($\tau_p$ is the actual delay), $k_p=Nf_p/\Delta_f$ ($0\leq k_p\leq N-1$) is the normalized Doppler frequency ($f_p$ is the actual Doppler frequency), $\eta(m,n)$ is the noise, and  
\begin{equation}
\alpha_p(m,n)=
\begin{cases}
1, \   l_p \leq m \leq M-1 \\
\frac{N-1}{N}e^{-j2\pi \left(\frac{n-k_p}{N}\right)},\  0\leq m \leq l_p-1. \label{eq-OTFS}
\end{cases}
\end{equation}
For sensing with the OTFS signal, the delay resolution is $1/(M\Delta_f)$, and the Doppler resolution is $1/(NT)$, where $T=MT_s$.

\section{Range and speed detections, and interference cancellations }

In \cite{IEEE:FAOR}, we proposed a fast algorithm for OTFS radar (FAOR) that enables efficient range and speed detection. However, this method cannot be directly implemented on our developed JRC prototype. In the JRC system, we utilize the NI Universal Software Radio Peripheral (USRP) for the transmission and reception of OTFS signals. Due to the close proximity of the transmit and receive antennas, significant leakage occurs from the transmitted signal into the received signal, resulting in strong interference between the transmitted and reflected signals. This self-interference can be powerful enough to obscure the reflected signals, rendering radar targets undetectable.

To address this challenge, we propose an interference cancellation scheme specifically designed to improve radar target detection. In this approach, self-interference is treated as a radar target with zero delay, which is then cancelled using the proposed method. The enhanced FAOR algorithm, which incorporates self-interference cancellation, is summarized as follows:
\begin{itemize}
\item \textbf{2D FFT Processing}: Perform the 2D-FFT on the received signal and the transmitted signal on DD domain to obtain ${Y}(a,b)$ and ${X}(a,b)$. Subsequently compute ${D}(a,b)={Y}(a,b) {X}^*(a,b)$.

\item
\textbf{Self-Interference Cancellation}: Averaging ${D}(a,b)$ across all data subcarriers to quantify the contribution of this zero-delay radar target, which represents the self-interference. This allows for effective cancellation of the self-interference, thereby enhancing the detection of actual radar targets as follows:
\begin{equation}
\tilde{{D}}(a,b)={D}(a,b)-\frac{1}{|S_{data}|}\sum_{a \in S_{data}} {D}(a,b)\label{eq-eFAOR3}
\end{equation}
where $S_{data}$ is the set of data subcarriers and $|S_{data}|$ is the cardinality of $S_{data}$.

\item \textbf{Range Doppler Map (RDM) Computation}: Compute the range Doppler map (RDM) after self-interference cancellation using the 2D IFFT as follows:
\begin{align}
&{R}(m,n)=\frac{1}{\sqrt{MN}}\sum_{a=0}^{M-1}\sum_{b=0}^{N-1} \tilde{D}(a,b) \nonumber\\
&\cdot \exp\left(j\frac{2\pi ma}{M}\right)  \exp\left(j\frac{2\pi bn}{N}\right).\label{eq-eFAOR4}
\end{align}

\item \textbf{Target Detection}: Determine the normalized range $l_1$ and speed $v_1$ of the targets by identifying the peaks in the RDM $|{R}(m,n)|$:
\begin{equation}
(l_1,k_1)=\arg \max_{m,n}|{R}(m,n)|.\label{eq-eFAOR5}
\end{equation}

\end{itemize}

\section{Breathing rate and heartbeat detection }

When the target is a human, the JRC prototype is capable of detecting both breathing rate and heartbeat.
In this section, we refine the system model to account for the effects of breathing and heartbeat. The transmitted OTFS RF signal is expressed as: \
$ a(t)=\text{Re}\left\{s(t)e^{j2\pi f_c t}\right\} $, where $s(t)$ denotes the baseband signal and
$f_c$ is the center frequency. When this signal
$a(t)$ reflects off a human target, which experiences time-varying chest-wall motion
$x(t)$ at a distance $r$, the total distance traveled between the transmitter and receiver becomes
$2r + 2x(t)$. Consequently, the round-trip time can be defined as $2d(t) = \frac{2r}{c}+ \frac{2x(t)}{c}$, where $c$ represents the speed of light. Therefore, the received signal at the radar receiver can be expressed as
\begin{equation}
g(t)=h \cdot \text{Re}\left\{s(t-2d(t))e^{j2\pi f_c(t-2d(t))}\right\}
\end{equation} where $h$ denotes the complex channel gain. The received signal is then downconverted to baseband as
$\tilde{g}(t)=h \cdot s(t-2d(t))e^{-j4\pi f_cd(t)}.$

Given the relatively slow movement of the chest wall due to breathing and the heartbeat, the delay remains an approximately constant value over a short radar detection interval. This stability enables us to concentrate on identifying longer-term variations for detection of vital signs \cite{IEEE:vitasign}. By continuously monitoring the phase and amplitude fluctuations in the radar returns, we can extract the periodic components corresponding to the breathing rate and heartbeat. The details are shown as follows. 
 \begin{itemize}
\item \textbf{Collecting Multiple Radar Detections:}  Let the radar detection duration be $\Delta$, which includes both the duration of the data frame and the idle time. Let $l=0,1,…,L_1-1$ represent the index for each detection interval. The RDM at the $l$-th detection can be expressed as
\begin{equation}
\tilde{{R}}(m,n,l)={R}(m,n)e^{-j2 \pi f_c \left(\frac{2r}{c}+\frac{2x(l\Delta)}{c} \right)}.
\end{equation}

\item \textbf{Analyzing Phase Variations:}
We identify the peak in the RDM for each radar detection, resulting in $ \rho(l)={{R}}(l_1,k_1)e^{-j2 \pi f_c \left(\frac{2r}{c}+\frac{2x(l\Delta)}{c} \right)}.$
The phase of each peak is $\phi(l)=- f_c \left(\frac{2r}{c}+\frac{2x(l\Delta)}{c}\right)+\theta.$ Clearly, $\phi(l),\ l=0,1,…,L_1-1$, can be utilized to estimate the frequency (change rate) of the slow motion of the chest $x(t)$, specifically pertaining to breathing and heartbeat rates.

\item \textbf{Filtering and spectral analysis}: We use bandpass filters to isolate the frequency components associated with the breathing and heartbeat signals, and then find the peak of the signal spectrum, as shown below:
\begin{equation}
\hat{b}_{br}=\max_{k\in \{0,1,2,...,L-1\}}\frac{1}{L}\sum_{l=0}^{L-1} {H}_{br}(\phi(l))e^{-\frac{j2\pi lk}{L}}
\end{equation}
\begin{equation}
\hat{b}_{hb}=\max_{k\in \{0,1,2,...,L-1\}}\frac{1}{L}\sum_{l=0}^{L-1} {H}_{hb}(\phi(l))e^{-\frac{j2\pi lk}{L}}
\end{equation}
where ${H}_{br}(\phi(l))$ and ${H}_{hb}(\phi(l))$ are the breathing and heartbeat signal after bandpass filtering, respectively, and $L\ge L_1$ is the FFT size.

\item
\textbf{Calculating Vital Signs}: The breathing rate
$f_{br}$ and heartbeat rate
$f_{hb}$ are estimated as:
\begin{equation}
f_{br}=\frac{\hat{b}_{br}}{L\Delta}, \
f_{hb}=\frac{\hat{b}_{hb}}{L\Delta}.
\end{equation}
The process is shown in Fig.~\ref{fig:vital-sign}.
\vspace{-3mm}
      \begin{figure}[h!]
        \centering
        \includegraphics[width=7.5cm,angle=0]{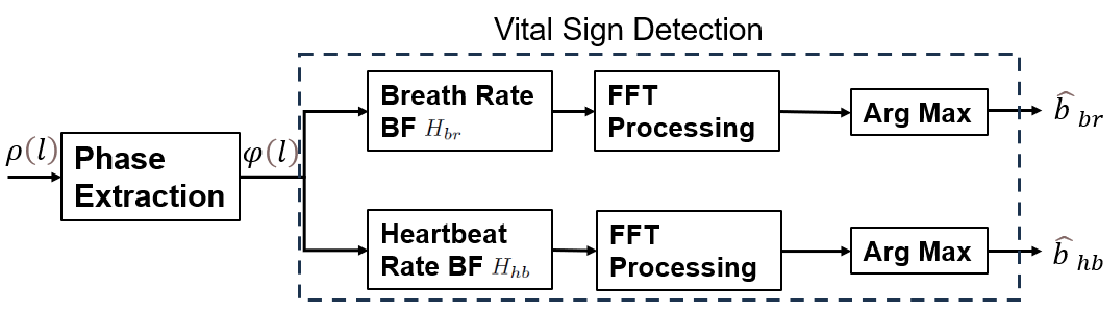}

\vspace{-2mm}
        \caption{Vital sign detection block}
        \label{fig:vital-sign}
\end{figure}

\end{itemize}

\section{Human Target Detection}

\begin{figure}[H]       
      \centering
        \includegraphics[width=8cm,angle=0]{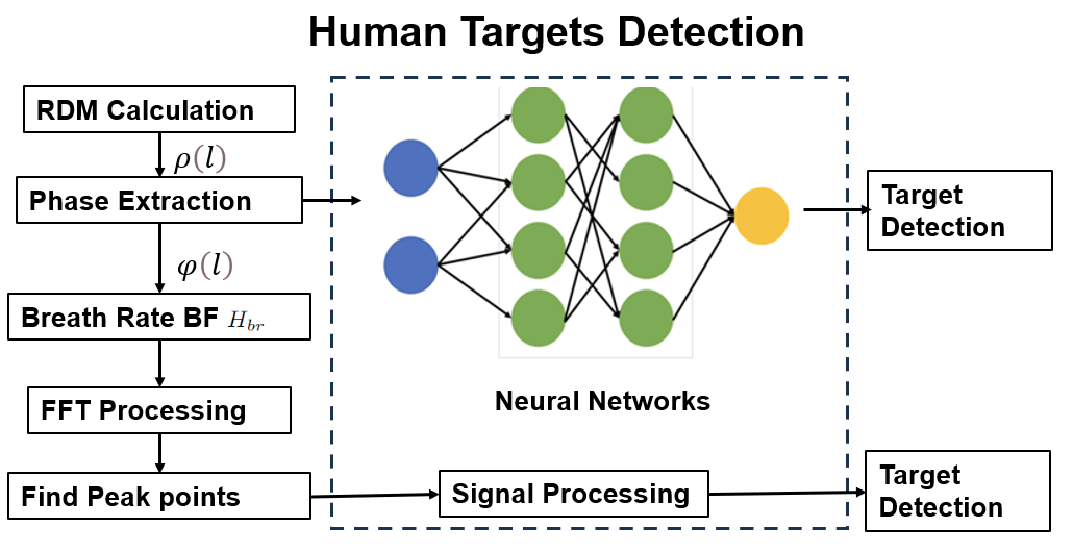}
        \vspace{-4mm}
        \caption{Human target detection procedure}
        \label{fig:human}
\end{figure}

In addition to range, speed, and vital sign detection, we integrated human target detection functionality into our prototype. As shown in Fig.~\ref{fig:human}, we proposed two methods to differentiate between human and non-human targets. The first method leverages machine learning: after extracting phase variations from the peaks of Range-Doppler Maps (RDMs), these variations are input into a neural network to distinguish between human and non-human targets. The second method applies signal processing to the extracted phase variations, following a breathing rate filter.
        \vspace{-4mm}

\section{Prototype setup and evaluation results}

\begin{figure}[h!]
        \centering
        \includegraphics[width=8cm,angle=0]{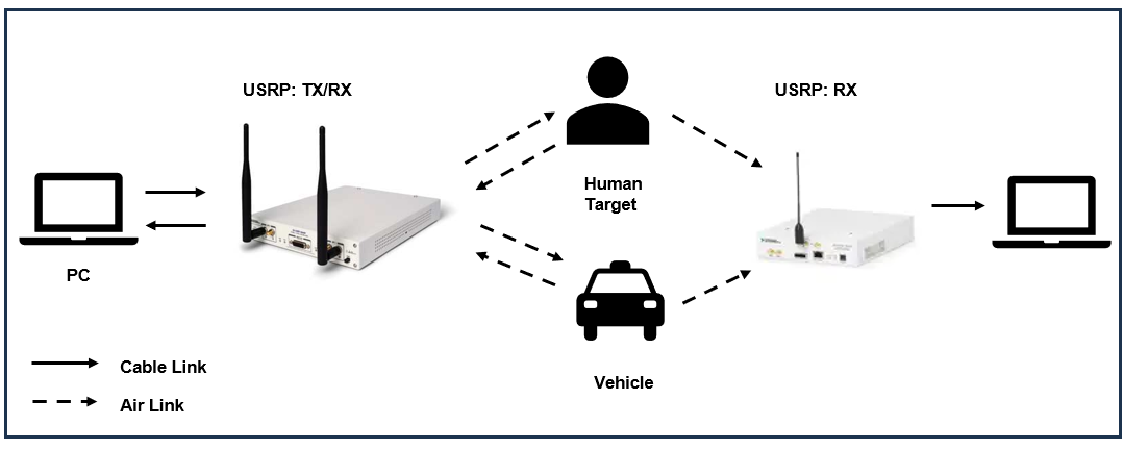}
        \vspace{-3mm}
        \caption{JRC prototype diagram}
        \label{fig:prototype}
\end{figure}

Fig.~\ref{fig:prototype} presents the block diagram of the developed JRC prototype, which comprises two primary components: the Universal Software Radio Peripheral (USRP) for transmitting and/or receiving OTFS signals, and PCs for signal generation and post-processing. This configuration facilitates integrated sensing and communication capabilities, allowing for the detection of range, speed, and vital signs, as well as the ability to distinguish between human and non-human targets. In this setup, the PC on the left is responsible for radar detection and processes the reflected signals received from the USRP on the same side. Conversely, the PC on the right handles data decoding after processing the received signals from the corresponding USRP.

%

\begin{figure}[h!]
        \centering
        \includegraphics[width=5cm,angle=0]{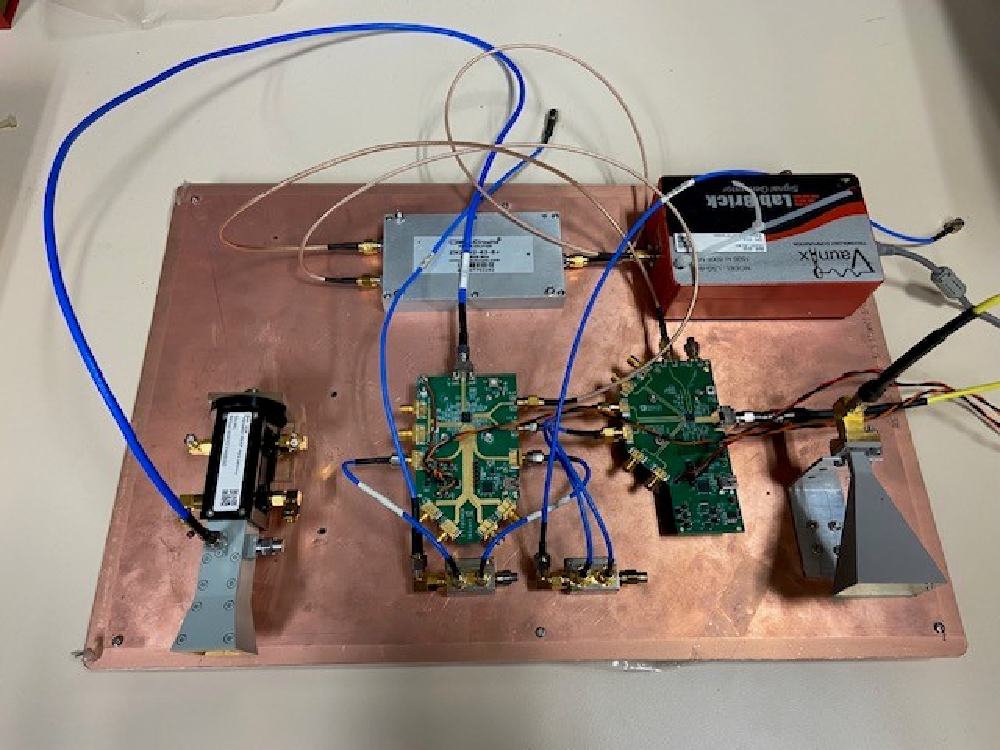}
        \vspace{-3mm}
        \caption{Physical setup of the mmWave JRC prototype}
        \label{fig:setup}
\end{figure}

Fig.~\ref{fig:setup} illustrates the hardware configuration of the developed JRC prototype. The transmitted OTFS signal is generated using MATLAB code, while the NI USRP2944 is employed for both transmitting and receiving OTFS signals. To manage mmWave frequencies between 24 GHz and 44 GHz, we utilize the EVAL-ADMV1013 and EVAL-ADMV1014 evaluation boards to up-convert sub-6 GHz signals to mmWave frequencies and vice versa.


The parameters we use for evaluation are as follows: Channel bandwidth: 100 MHz, Carrier frequency: 29 GHz, Subcarrier spacing: 30 kHz, Symbols per slot: 14,
Slots per subframe: 1, Slots per frame: 20, Number of transmit antennas: 1, Number of receive antennas: 1.


      \begin{figure}[h!]
        \centering
        \includegraphics[width=8cm,angle=0]{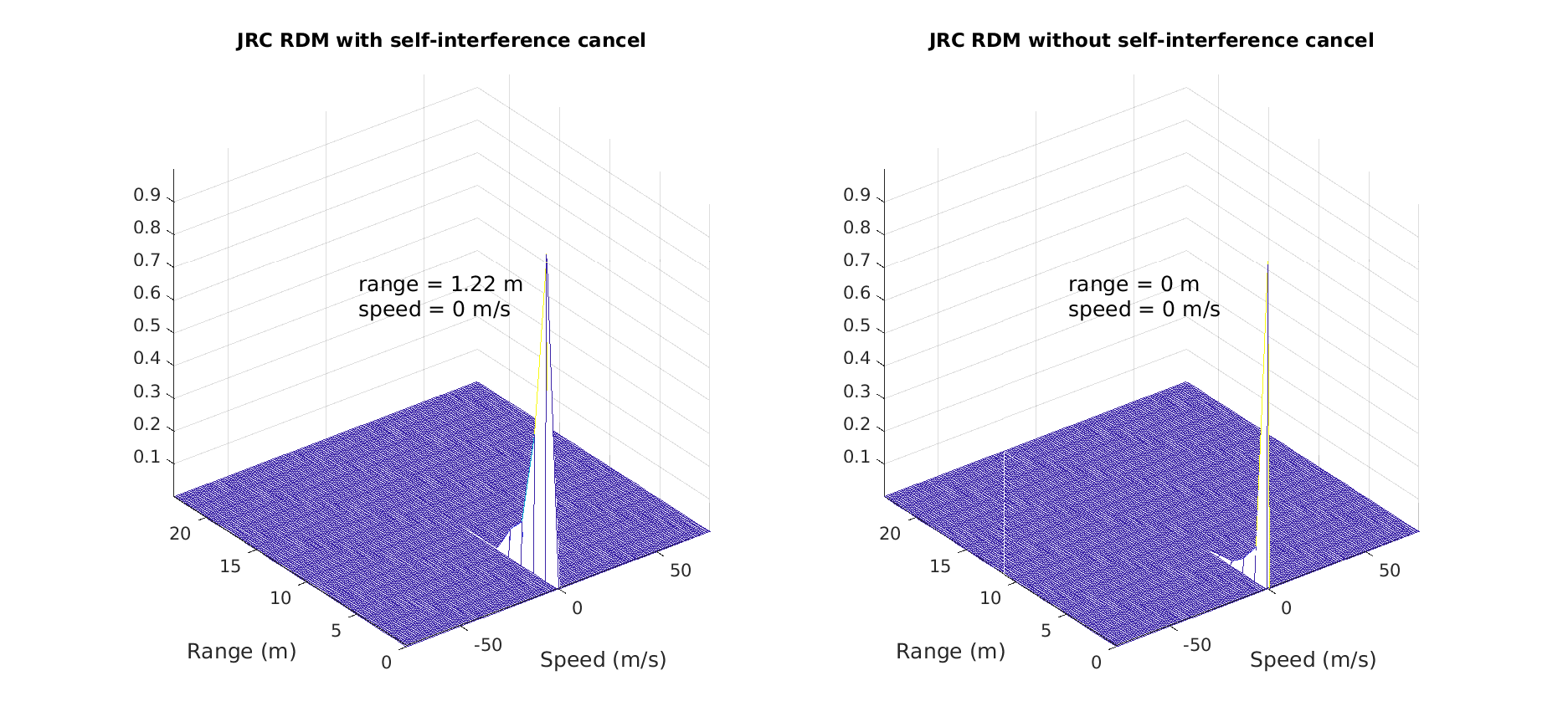}
        \vspace{-3mm}
        \caption{Range and speed detection performance}
        \label{fig:performance1}
        \end{figure}

      \begin{figure}[h!]
        \centering
        \includegraphics[width=5.5cm,angle=0]{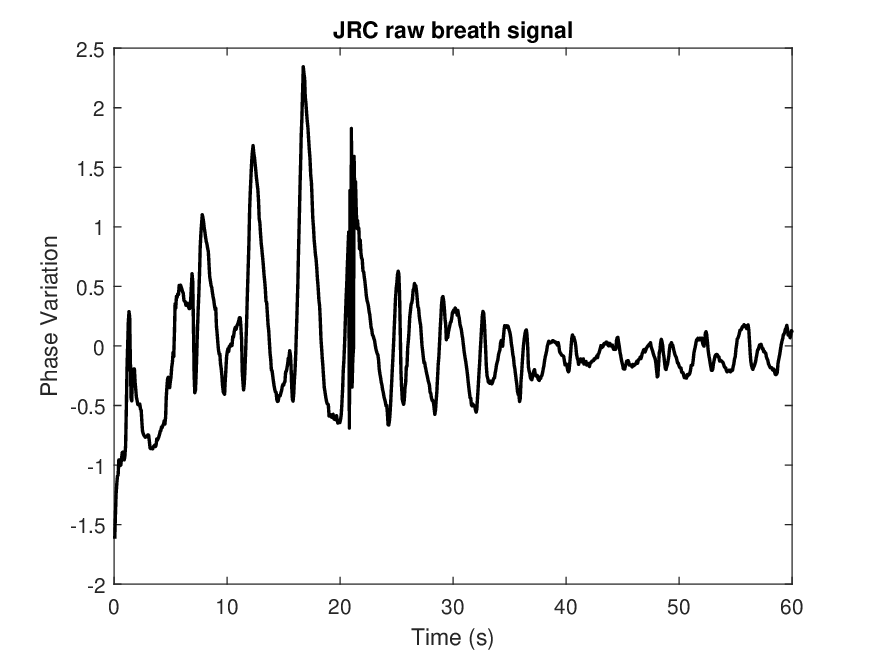}

\vspace{-4mm}
        \caption{Phase variations with time}
                \label{fig:performance2}
        \end{figure}
      \begin{figure}[h!]
        \centering
        \includegraphics[width=5.5cm,angle=0]{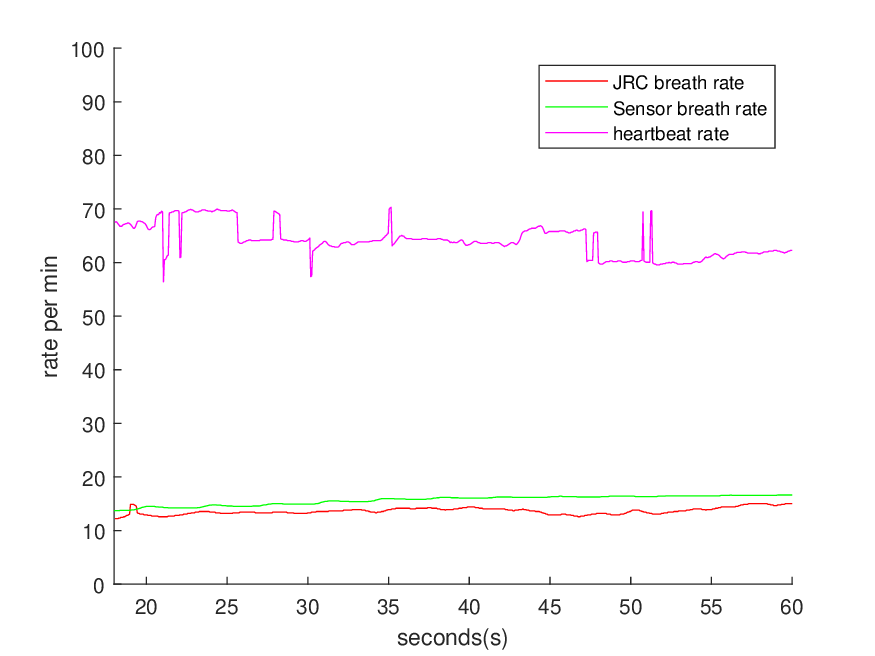}

\vspace{-4mm}
        \caption{Vital sign detection performance}
        \label{fig:performance3}
 \end{figure}

%

      \begin{figure}[h!]
        \centering
        \includegraphics[width=5.5cm,angle=0]{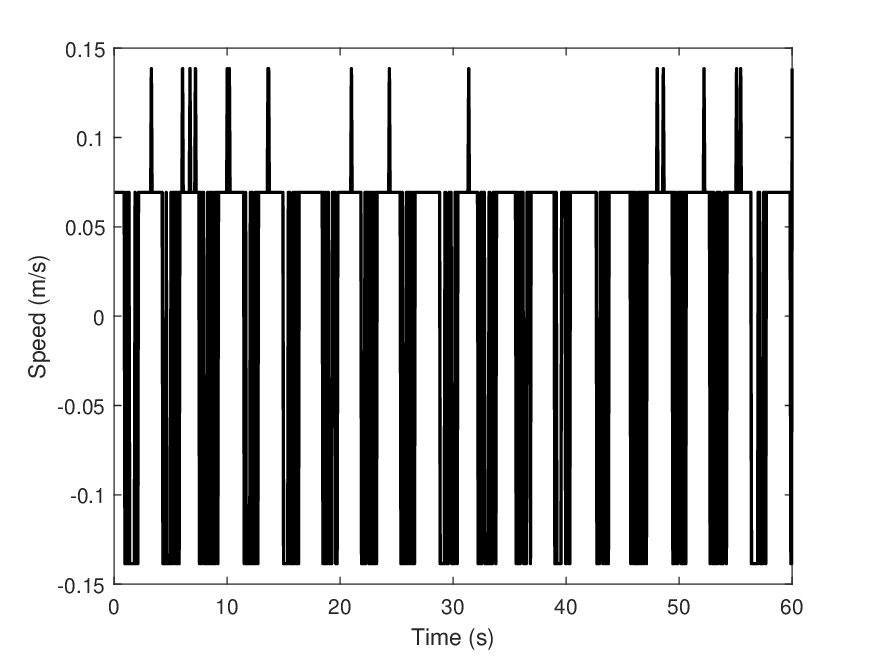}

\vspace{-4mm}
        \caption{Speed detection result for a walking human}
        \label{fig:performance4}
\end{figure}



We analyze two scenarios for range, speed, and vital sign detection. In the first scenario, we focus on static targets, specifically a human seated approximately 1.2 meters away. Fig.~\ref{fig:performance1} compares the results of range and speed detection with and without self-interference. In the absence of self-interference cancellation, self-interference can obscure the real target, rendering it undetectable. However, with the proposed self-interference algorithm, we can successfully diminish self-interference and accurately detect the target. Fig.~\ref{fig:performance2} illustrates the phase variations of the peaks in the Range-Doppler Map (RDM), collected at radar detection intervals of 0.06 seconds, which reflect the movements of the human chest over time, providing information on the breathing rate and heartbeat.


To benchmark our vital sign detection, the individual wore a respiration sensor (SA9311M), which provides breathing rate data based on physical chest movements. Fig.~\ref{fig:performance3} indicates that the breathing rate detected by the JRC prototype closely aligns with the measurements obtained from the respiration sensor. However, the detection of the heartbeat displays fluctuations due to various interfering frequencies, which may stem from human movement, harmonics of breathing signals, or reflected noise from the surrounding environment. Increasing the frequency of the OTFS signals could potentially enhance the detectability and reliability of the heartbeat signals.

In the second scenario, we investigate moving targets, where either a human or a robot moves back and forth in front of the prototype. Fig.~\ref{fig:performance4} presents the speed results for a moving human, demonstrating a high speed resolution capability of up to 0.1 m/s. The positive and negative speeds depicted in the figure correspond to the forward and backward movements of the targets.


For human detection with our prototype, we collected data from 10,000 cases, evenly divided between 5,000 human cases and 5,000 non-human cases. The signal processing-based model correctly identified all 5,000 human cases, and out of the 5,000 non-human cases, it accurately identified 4,359. In the machine learning-based approach, we employed a 1D convolutional neural network (CNN). The architecture consists of three convolutional layers with 64, 128, and 256 filters for feature extraction, followed by max pooling layers. A flatten layer converts the 2D feature maps into a 1D vector, which is then processed by a dense layer with 32 units. The output layer uses a sigmoid activation function for probability outputs. The results for the training data were as follows:

\begin{itemize}
\item Out of 3,626 non-human cases, the model correctly identified all 3,626.
\item Out of 3,626 human cases, it correctly identified 3,607.
\end{itemize}
For the test data:

\begin{itemize}
\item Out of 1,506 non-human cases, the model correctly identified all 1,506.
\item Out of 1,540 human cases, it correctly identified 1,532.
\end{itemize}
These results demonstrate strong performance for both methods, with near-perfect accuracy in the machine learning approach across both training and test datasets.

\section{Conclusions}
We have developed a JRC system and prototype based on OTFS signals, incorporating a fast radar sensing algorithm for detecting target range and speed with self-interference cancellation. Additionally, we implemented practical methods for detecting human vital signs, such as breathing rate and heartbeat, following target detection. To distinguish between human and non-human targets, we proposed two approaches: one based on signal processing and the other on machine learning. Experimental results have shown the effectiveness of the prototype in detecting range, speed, and vital signs in both human and mobile robot scenarios, as well as in distinguishing between human and non-human targets. The prototype serves as a versatile platform for evaluating various JRC algorithms, offering valuable insights for the development of practical applications.

\end{document}